\documentclass[a4paper,11pt]{article}
\usepackage{pos}

\usepackage{tikz}
\usepackage{tikz-feynman}
\tikzfeynmanset{compat=1.1.0} 

\tikzfeynmanset{
    doublefermion/.style={
        /tikz/draw=black,
        double distance=1.2pt, 
        line width=0.4pt,
        postaction={draw, line width=0.4pt, white, shorten >=0.4pt, shorten <=0.4pt}
    }
}
\usepackage[export]{adjustbox}

\renewcommand{\logo}{\relax} 

\def\au{a_\mu}
\def\ae{a_e}
\def\aue{a_{\mu-e}}

\def\auehvp{a_{\mu-e}^\mathrm{HVP}}
\def\auhvplo{a_\mu^\mathrm{HVP,LO}}

\def\auehvplo{a_{\mu-e}^\mathrm{HVP,LO}}

\def\ga{\left(}
\def\dr{\right)}
\usepackage{physics}

\def\eg{\emph{e.g.}}

\title{Short-Distance-Free Leptonic $g-2$}

\author*[a]{Siyuan Li}
\author[a]{Vladimir Pascalutsa}
\author[b,c]{Maxim Pospelov}

\affiliation[a]{Institut f\"ur Kernphysik, Johannes Gutenberg-Universit\"at  Mainz, \\ D-55128 Mainz, Germany
}

\affiliation[b]{William I.\ Fine Theoretical Physics Institute, University of Minnesota,\\
MN 55455, United States}
\affiliation[c]{Theoretical Physics Department, CERN,\\
1 Esplanade des Particules, CH-1211 Geneva 23, Switzerland}

\emailAdd{siyuan.li@uni-mainz.de}

\abstract{
We note that the hadronic contributions to $a_e$ and $a_\mu$ are strongly correlated, which potentially allows one to reduce the uncertainties associated with higher-energy contributions (and address the current discrepancy between lattice and data-driven evaluations) by improved measurements of $a_e$ and the fine-structure constant.
 }

\FullConference{14th International Workshop on e$^+$e$^-$ collisions from Phi to Psi 2026 (PHIPSI26)\\
8-11 June 2026\\
Pisa, Italy\\}

\begin{document}
\maketitle

\section{Introduction}

The anomalous magnetic moment of a lepton (electron, muon and tau), $a_\ell = (g-2)_\ell/2$, has attracted extensive study in both theoretical predictions and experimental measurements, partially due to its potential in probing physics beyond the Standard Model (BSM)~\cite{Aliberti:2025beg,Fan:2022eto,Aguillard:2025fij,Melnikov:2006sr,Jegerlehner:2017gek,Aoyama:2020ynm}. 
One of the most prominent experimental results was published by Fermilab on the muon $(g-2)$, achieving unprecedentedly high precision on the observable~\cite{Aguillard:2025fij}.

Among all contributions to the theoretical evaluation of $a_\mu$, the leading-order (LO) hadronic vacuum polarization (HVP) contribution, $\auhvplo$, is the dominant source of uncertainty. Furthermore, the determinations by data-driven dispersive approach and lattice QCD exhibit a persistent tension (see reviews in \eg~\cite{Aliberti:2025beg,Aoyama:2020ynm}). While lattice QCD provides robust results that align with experiment, there is still room for improvement in systematic effects such as isospin-breaking effects and discretization errors.

In our work, we aim to exploit the strong correlation between muon and electron $(g-2)$ and combine them in such a manner that many systematic effects cancel out. 
The method utilizes the fact that the electron mass, $m_e$, is negligible compared to the hadronic scale $\Lambda_{\rm QCD}$. The electron therefore does not resolve hadronic structure: $a_e^{\rm HVP}$ is simply $m_e^2$ times the slope of the vacuum polarization at zero momentum. Since contributions from scales well above $m_\mu$ scale as $m_\ell^2$ for both leptons, $a_e$ can be used to cancel the high-energy part of $a_\mu$ after proper rescaling:
\begin{equation}\label{eq:aue_def}
    \aue\equiv \au - (m_\mu/m_e)^2\ae,
\end{equation}
which is a well-defined, measurable observable that isolates the sub-GeV range of hadronic contributions of $\au$. With future improvements in $a_e$ and the fine structure constant $\alpha$, $\aue$ would be the only remaining piece to compute for $\au$ determination.

The purpose of $\aue$ is not only to reduce the discrepancy in the $\au$ determination between lattice QCD and data-driven approaches, but also among different data-driven evaluations themselves. To demonstrate the advantage of our new observable, Table~\ref{tab:ref_values} presents the current HVP LO values of $\au$, $\ae$ and the corresponding $\aue$. We assume $100\%$ correlation between $\au^\mathrm{had}$ and $\ae^\mathrm{had}$ and find that both value and uncertainty are significantly reduced in $\auehvplo$ compared to the muon.
\begin{table}[hbt]
\centering
\renewcommand{\arraystretch}{1.5}
\begin{tabular}{c|llr}
\hline\hline
LO HVP & $\au\ga 10^{-10}\dr$ & $\ae\ga10^{-14}\dr$ & $\aue \ga 10^{-10}\dr$\\
 \hline
lattice QCD       & $713.2(6.1) $~\cite{Aliberti:2025beg}   & $189.3 (8.2)$~\cite{Budapest-Marseille-Wuppertal:2017okr} & $-96.1(29.0)$
\\
data-driven    &  $692.8 (2.4)$~\cite{Keshavarzi:2019abf}  & $186.10(66)$~\cite{Keshavarzi:2019abf}
& $-102.8(4)$\\
\hline 
discrepancy    & $20.4(6.6)$   &$3.2(8.2)$ & $6.7(29.0)$\\
\hline\hline
\end{tabular} 
\caption{LO HVP contribution to $a_\mu,\, a_e$, and $\aue$ from lattice QCD and data-driven approaches.}
\label{tab:ref_values}
\end{table}

In this report, based on Ref.~\cite{Li:2026bgj}, we provide a direct calculation of $\auehvplo$ in Sec.~\ref{sec:kernel_fcn_and_channel} and discuss the application of chiral perturbation theory ($\chi$PT) to $\aue$. We also provide an analysis of the potential of $\aue$ as a probe of BSM physics, especially its high sensitivity in the sub-GeV region in Sec.~\ref{sec:BSM}. 

\section{Kernel Function and Channel Contribution}\label{sec:kernel_fcn_and_channel}
The HVP contribution to the leptonic $(g-2)$ is give by the dispersion relation~\cite{Bouchiat:1961lbg,PhysRev.168.1620,Lautrup:1968tdb,Gourdin:1969dm}
\begin{equation}
    a_\ell^{\mathrm{HVP}} = \frac{\alpha}{\pi^2}\int\limits_{s_0}^\infty \dd s \, \frac{\mathrm{Im} \Pi(s)}{s} 
K_\ell(s) \,,
\end{equation}
where $\Pi (s)$ is the HVP function and $K_\ell (s)$ is the kernel function with lepton mass $m_\ell$,
\begin{equation}
    \label{eq:Ku}
    K_\ell(s) = \frac{m_\ell^2}{s} \int\limits_0^1 \dd x \frac{ x^2 (1-x)}{1-x+x^2( m_\ell^2/s)}.
\end{equation}
At the hadronic scale, the lower limit of the energy scale $s$ is associated with the pion-production threshold. Therefore, for the electron, we can take the limit $m_e^2/s\rightarrow0$ and find
\begin{equation}\label{eq:Ke}
    K_e (s) \simeq m_e ^2/3s,
\end{equation}
yielding the kernel function for $\aue$ 
\begin{equation}\label{eq:Kue}
    K_{\mu-e}(s) =  K_\mu(s) - \frac{m_\mu^2}{m_e^2} K_e(s) \simeq 
-\frac{m_\mu^4}{s^2}
\int\limits_0^1 \dd x 
\frac{ x^4}
{1-x+x^2 (m_\mu^2/s)}.
\end{equation}
Eq.~\eqref{eq:Ke} shows explicitly that $K_\ell \propto m_\ell^2$ at high-energy scale and this term is expected to cancel in $\aue$~\eqref{eq:aue_def}.
The high-energy suppression can also be argued from the extra power of $1/s$ in the $(\mu-e)$ kernel~\eqref{eq:Kue} compared to that of individual leptons~\eqref{eq:Ku}.
An analogous argument can be made for lattice QCD in the time-momentum representation as well~\cite{Bernecker:2011gh}, where the short-time contribution is suppressed in $\aue$ resulting in reduced sensitivity to systematic effects, such as discretization errors, compared to $a_\ell$~\cite{Beltran:2026ofp}.

One can then evaluate the dispersion integral channel by channel, \eg~starting with the leading $\pi^+\pi^-$-channel. We adopt the vector-meson-dominance (VMD) type model (see \eg ~Ref.~\cite{Biloshytskyi:2025fjm}), include isospin-breaking effects, and compare our results with the data-driven literature results from Ref.~\cite{Keshavarzi:2019abf} in Table~\ref{tab:pipi_breakdown}, which shows qualitative agreement. 
\begin{table}[htb]
\renewcommand{\thempfootnote}{\alph{mpfootnote}} 
\renewcommand{\arraystretch}{2}
\centering
\begin{tabular}{c|lll}
\hline\hline
 Source & $\au^{\pi^+\pi^-} (10^{-10})$  & $\ae^{\pi^+\pi^-} (10^{-14})$ & $\aue^{\pi^+\pi^-} (10^{-10})$\\
 \hline
VMD phys~\cite{Biloshytskyi:2025fjm}    & $500.67$ & $137.42$ & $-86.84$ \\
KNT eval.~\cite{Keshavarzi:2019abf} &  $503.46(1.91)$ & $138.59(54)$ & $-89.1(4)$\\
\hline\hline
\end{tabular}  
\caption{The $\pi^+\pi^-$-channel contribution in a VMD model compared with a current data-driven evaluation.}
\label{tab:pipi_breakdown}
\end{table}

\begin{figure}
    \centering
    \includegraphics[width=0.5\linewidth]{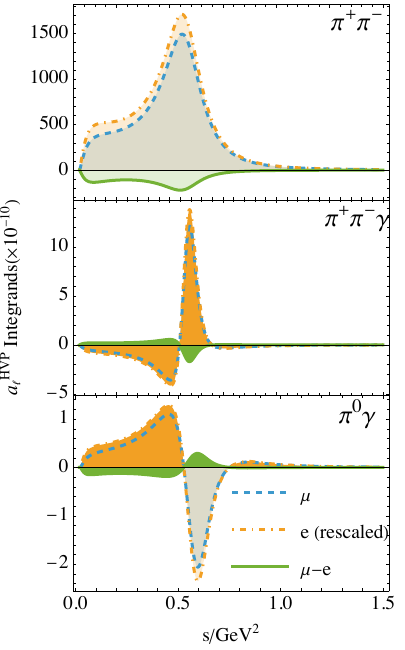}
    \caption{The integrand 
    of $a_\ell (\times 10^{-10})$ at $\pi^+ \pi^-$, $\pi^+ \pi^-\gamma$ and $\pi^0\gamma$ channels for muon, rescaled electron and their difference $\mu-e$.}
    \label{fig:integrands}
\end{figure}

The integrands of the corresponding observables in Table~\ref{tab:pipi_breakdown} are illustrated in Fig.~\ref{fig:integrands}, along with the same analysis for the subleading channels from $\pi^+\pi^-\gamma$ and $\pi^0\gamma$. The plot shows large cancellation between the muon and rescaled-electron contributions in the high-energy region, while sub-GeV contributions, such as the $\rho$-meson resonance, survive in $\aue$. 

\begin{figure}[htb]
\centering
\begin{tikzpicture}[scale=1]
\begin{feynman}

\vertex (a) at (-3.2,1.0){\(\gamma\)};
\node[dot, minimum size=1mm] (b) at (-2.67,0.15);
\node[dot, minimum size=1mm] (c) at (-2,-0.2);
\vertex (d) at (-4,-1){\(\ell\)};
\node[dot, minimum size=1mm] (e) at (-3.5,-1);
\node[dot, minimum size=1mm] (f) at (-2.5,-1);
\node[dot, minimum size=1mm] (g) at (-1.5,-1);
\vertex (h) at (-1,-1){\(\ell\)};

\node[dot, minimum size=1mm] (a2) at (0.55,-0.5);
\node[dot, minimum size=1mm] (b2) at (1.1,-0.5);
\node[dot, minimum size=1mm] (c2) at (1.9,-0.5);
\node[dot, minimum size=1mm] (d2) at (2.45,-0.5);
\node[dot, minimum size=1mm] (e2) at (1.5,0.5);
\vertex (f2) at (0.1,-1){\(\ell\)};
\vertex (g2) at (2.9,-1){\(\ell\)};
\vertex (h2) at (1.5,1.2){\(\gamma\)};
\diagram*{
  (a) -- [photon] (b)
      -- [scalar, edge label=\(\pi^0\)] (c)
      -- [photon] (f),
  (c) -- [photon] (g),
  (b) -- [photon] (e),
  (d) -- [plain] (h),

  (a2) -- [photon] (b2)
       -- [photon, half left] (c2)
       -- [photon] (d2),
  (c2) -- [scalar, half left, edge label=\(\pi^0\)] (b2),
  (h2) -- [photon] (e2),
  (f2) -- [plain] (e2) -- [plain] (g2),
};

\end{feynman}
\end{tikzpicture}
    \caption{Pseudoscalar-meson contribution to light-by-light (left) and isospin-breaking HVP (right).}
    \label{fig:lbl_pi0gamma_feynman_diagrams}
\end{figure}
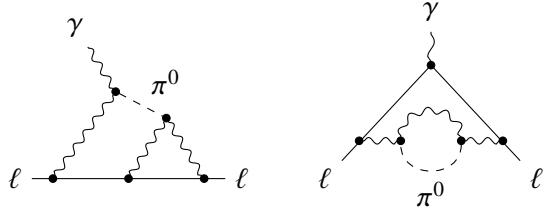

\begin{figure}[htb]
    \centering
    \includegraphics[width=0.5\linewidth]{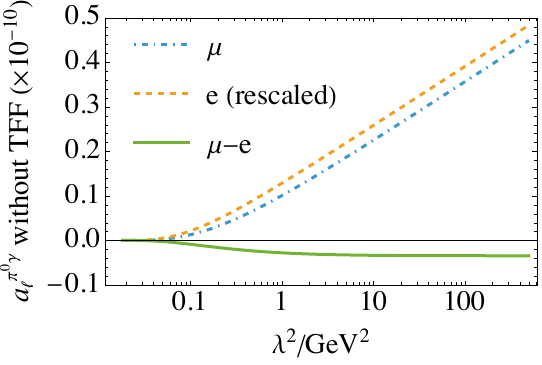}
    \caption{With $\lambda^2$ being the cut-off of the dispersion relation integral, $\au^{\pi^0\gamma}$ and $\ae^{\pi^0\gamma}$ (rescaled) diverge without the implementation of TFF while $\aue^{\pi^0\gamma}$ converges since the UV divergences cancel exactly in the rescaled subtraction.}
    \label{fig:convergence}
\end{figure}

The $\pi^0\gamma$ channel includes the $\pi\gamma\gamma$ vertices (Fig.~\ref{fig:lbl_pi0gamma_feynman_diagrams}) which introduce ultraviolet (UV) divergences. This UV divergence is typically regulated by an empirical pion transition form factor (TFF) at the $\pi\gamma\gamma$ vertex, along with the disadvantages it introduces, such as short-distance constraints~\cite{Melnikov:2003xd,Bijnens:2019ghy,Colangelo:2019lpu}. However, the subtracted observable $\aue$ cancels the UV divergence as shown in Fig.~\ref{fig:convergence}, allowing one to evaluate the $\pi^0\gamma$ channel directly with $\chi$PT. 
The EFT result of $\aue^{\pi^0\gamma} = -0.34(5)\times 10^{-11}$ does not show good agreement with TFF-dependent value of $-0.45\times 10^{-11}$~\cite{Biloshytskyi:2025fjm}. We aim to extend the UV convergence benefit of $\aue$ to the two-loop hadronic light-by-light (LbL) contribution which has the same $\pi\gamma\gamma$ vertices (Fig.~\ref{fig:lbl_pi0gamma_feynman_diagrams}) in our future work.
The pointlike $\pi\gamma\gamma$
vertices in the LbL contributions have the leading UV-divergent behaviour like~\cite{Knecht:2001qg,Ramsey-Musolf:2002gmi}
\begin{equation}\label{aell_lbl}
a_\ell^{\pi^0,\,\mathrm{LbL}} = \ga\frac{\alpha}{\pi}\dr^{3}
\frac{3 m_\ell^{2}}{16\pi^{2} f_\pi^{2}}\,
\ln^{2}\!\frac{\lambda}{m_{\pi^0}} + \dots,
\end{equation}
where $\lambda$ is the UV cutoff and 
the ellipsis stands for single-logarithmic and finite terms. The coefficient of
the double logarithm is the
same for muon and electron and 
multiplies $m_\ell^{2}$ by a function of $\lambda/m_{\pi^0}$ alone \cite{Marciano:2016yhf}, canceling in $\aue$. 

\section{BSM potential}\label{sec:BSM}
The lepton $(g-2)$ has long served as a probe of BSM effects. We argue that $\aue$ is more sensitive to new-physics models than $a_\ell$. 
A clear example of this advantage is the dark photon $A'$, a universally coupled new-physics particle with a mass scale well below the weak scale. The correction from the dark photon to $a_\ell$ enters at one-loop level~\cite{Fayet:2007ua,Pospelov:2008zw,Arkani-Hamed:2008kxc}
\begin{equation}
    a_\ell^{A'} = \frac{\alpha}{2\pi}\times \varepsilon^2 \int_0^1 \dd x \frac{2 m_\ell^2x^2(1-x)}{m_\ell^2x^2+m_{A'}^2(1-x)},
    \label{eq:aDP} 
\end{equation}
where $\varepsilon$ is the kinetic mixing parameter and $m_{A'}$ denotes dark photon mass. Eq.~\eqref{eq:aDP} is effectively equivalent to including a new delta-function term in the polarization operator, $\mathrm{Im} \Pi(s)\propto \varepsilon^2 \delta(s-m_{A'}^2)$, inside the dispersive formula. In the sub-GeV mass range of $m_{A'}$, the $(g-2)$ measurements can be incorporated into direct searches of $A'$~\cite{Fayet:2007ua,Pospelov:2008zw}.

\begin{figure}
    \centering
    \includegraphics[width=0.7\linewidth]{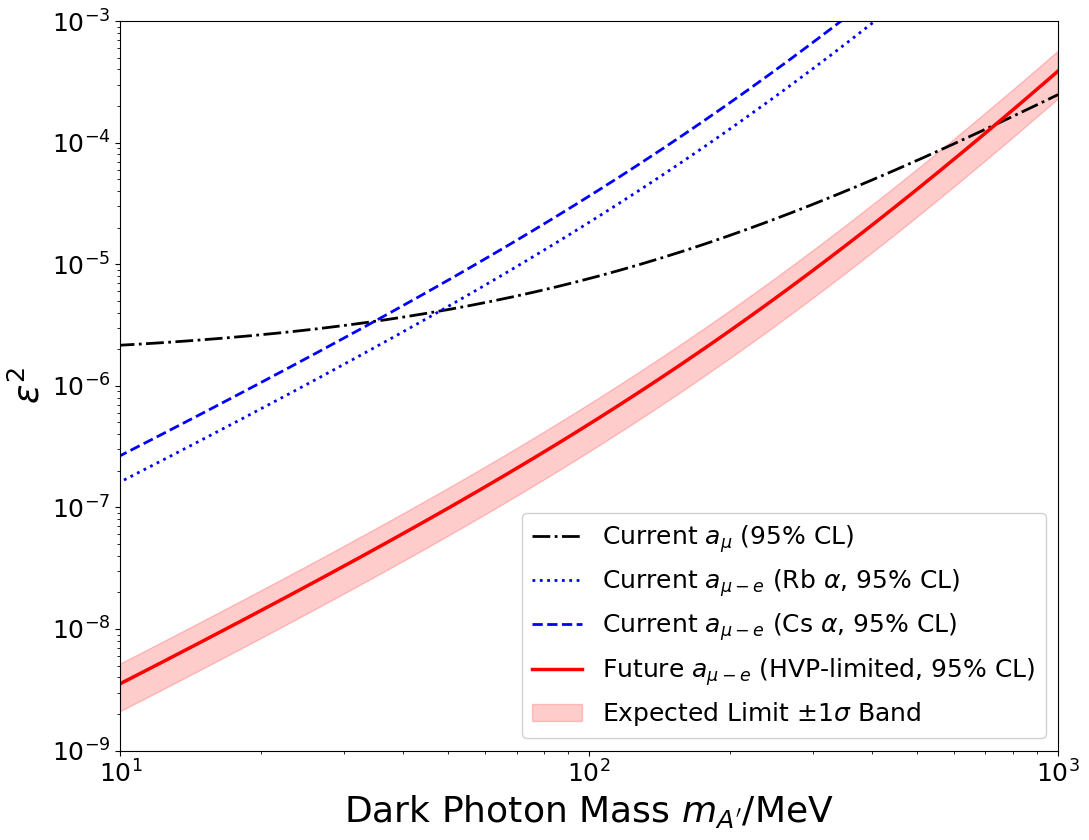}
    \caption{Exclusion plot for the dark-photon parameters. It is assumed that the only limitation comes from the HVP error, and 
    it is taken to be $\delta a_\mu^{\rm HVP}=1.0\times 10^{-9}$ (black) and $\delta a_{\mu-e}^{\rm HVP}=1.5\times 10^{-10}$ (red) for illustration. The blue lines represent the exclusions by $\auehvp$ with limitations from the current $\alpha$ measurements (Cs and Rb). Above colored lines the parameter space would be excluded. It is shown that $\aue$ would provide a deeper reach into $\varepsilon^2$ for all $m_{A'} \lesssim$ several hundred MeV.}
    \label{fig:darkphoton}
\end{figure}

Assuming that the experimental determinations of $a_e$ and the fine structure constant $\alpha$ have been improved to precision levels of $\delta\alpha/\alpha\sim 10^{-12}$ and $\delta a_e\sim 10^{-15}$, respectively, and that the only theoretical uncertainty comes from HVP contributions, we show that the sub-GeV dark photon parameter space from Eq.~\eqref{eq:aDP} could be better probed by $\aue$ in Fig.~\ref{fig:darkphoton}.

\section{Summary}\label{sec:summary}
We introduce $\aue = \au - (m_\mu^2/m_e^2) \, \ae$, a sub-GeV window observable that exploits the strong correlation between HVP contributions to the muon and electron anomalous magnetic moments at high energy. 
The formulation of $\aue$ can be written as a modification of the kernel function in the dispersion relation. $\auehvplo$ reduces the sensitivity in the HVP energy regime where lattice QCD and the data-driven approach show persistent tension and acts as a powerful probe of low-energy hadronic effects. 
Advantages of $\aue$ , such as being free of the UV divergence introduced by the $\pi\gamma\gamma$ coupling, allow an EFT calculation free of empirical form factors for the two-loop diagrams with pseudoscalar contributions. With future improvements of $\ae$ and $\alpha$~\cite{Mohr:2024kco,Fan:2022eto,Morel:2020dww,Parker:2018vye}, $\auehvp$ allows one to probe $\au$ without relying on the poorly constrained high-energy regime and acts as a powerful tool to probe deeper into BSM parameter space than individual leptonic anomalous magnetic moments.

\bibliographystyle{JHEP}
\bibliography{references.bib}



\end{document}